\documentclass[aps,pra,twocolumn]{revtex4}
\usepackage{latexsym}
\usepackage{bbm}
\usepackage{graphics}
\usepackage{here}
\usepackage{amsmath}
\usepackage{amssymb}
\usepackage{graphicx}

\newcommand{\beq}{\begin{equation}}
\newcommand{\eeq}{\end{equation}}
\newcommand{\bra}[1]{\langle #1 |}
\newcommand{\ket}[1]{| #1 \rangle}
\newcommand{\identity}{\hat{\openone}}
\newcommand{\icon}{{\cal C}_I}
\newcommand{\threecon}{{\cal C}_3}

\begin{document}

\title{Bipartite entanglement measure based on covariances}

\author{Isabel Sainz Abascal}
\affiliation{School of Information and Communication Technology,
Royal Institute of Technology (KTH), Electrum 229, SE-164 40 Kista,
Sweden}
\author{Gunnar Bj\"{o}rk}
\affiliation{School of Information and Communication Technology,
Royal Institute of Technology (KTH), Electrum 229, SE-164 40 Kista,
Sweden}

\date{\today}

\begin{abstract}
We propose an entanglement measure for two qu$N$its based on the
covariances of a set of generators of the {\it su}$(N)$ algebra. In
particular, we represent this measure in terms of the mutually
unbiased projectors for $N$ prime. For pure states this measure
quantify entanglement, we obtain an explicit expression which
relates it to the concurrence hierarchy, specifically the
$I$-concurrence and the 3-concurrence. For mixed states we propose a
separability criterion.
\end{abstract}

\maketitle

\section{Introduction}

Entanglement plays a key role in quantum information and quantum
communications processes. During the last years a wealth of
entanglement measures have been proposed and studied \cite{ref1}. In
particular, the two-qubit case has been extendedly studied, and
entanglement of formation \cite{formation} and concurrence
\cite{con} are now widely accepted as entanglement measures. These
measures require a complete knowledge of the density matrix, which,
in turn, require state tomography, an experimentally and
computationally labor intensive process.

Higher dimensional cases are more complicated. An accepted {\it
separability criterion} is the so-called positive partial transpose
(PPT) criterion \cite{ppt}, which is necessary and sufficient for
composite systems with dimensions $2 \times 2$ and $2 \times 3$
\cite{posmaps}, otherwise it is only necessary and it does not give
information about the amount of entanglement. Motivated by the
positive, but not completely positive maps, which are always
positive for separable states \cite{posmaps}, another important
criterion has been introduced \cite{reduction}. A separability
criteria which identifies the entanglement in some states that PPT
does not (so-called bound states), is the realignment method
\cite{realig}. The method has the advantage that it gives a rough
quantitative estimate of the degree of entanglement.

However, even for two qutrits there is no consensus on how to {\it
quantify} entanglement. Ulhmann introduced one measure that is based
on the fact that antilinear operators are nonlocal \cite{ulhmann}.
Unfortunately, this generalization is not invariant under local
unitary transformations, an important property that an entanglement
measure requires. Rungta {\it et al.} introduced another
generalization of concurrence \cite{concurrence}, namely the
$I$-concurrence, based on a generalization of the spin-flip
operation called universal inverter. The measure posses the
requirements for a good entanglement measure \cite{vedral}, and
theoretically is very nice, nevertheless the universal inverter is
not a complete positive operation, so that it is not directly
experimentally realizable. However, for bipartite systems with no
more than two eigenvalues different from zero there is an explicit
formula for the $I$-tangle, that is the square of the
$I$-concurrence \cite{osborne}. At roughly the same time, this
concurrence was also introduced in \cite{albeverio} in terms of
invariants under local unitary transformations.

For mixed states, the situation is further complicated. E.g., the
$I$-concurrence \cite{concurrence} requires a global minimization
over all bases which makes it cumbersome to calculate for mixed
states. Mintert, Ku\'s, and Buchleitner \cite{mintert} found a lower
bound on $I$-concurrence which is simpler to estimate than the
$I$-concurrence itself, and a short time later Chen, Albeverio and
Fei found an analytical lower bound \cite{mixcon} connecting the
$I$-concurrence with the PPT criterion \cite{ppt}, and the
realignment criterion \cite{realig}. Another attempt of generalizing
the concurrence \cite{con} for mixed states in higher dimensions was
made by Badziag \emph{et al.} in \cite{badziag}. They introduced the
so called pre-concurrence, which, unfortunately, is difficult to
analyze for states with a rank $>2$. Moreover, there is no guarantee
that the ensuing concurrence matrix can be diagonalized. Then, they
introduce the biconcurrence, which implies a separability criteria,
but it too, requires a minimization procedure. Yet another proposal
to deal with mixed states in higher dimensions is presented in
\cite{lee}, where the concept of negativity is extended
\cite{negativity} for mixed states by means of a convex-roof, which
gives a necessary and sufficient separability criteria. For two
qubits it coincides with the concurrence \cite{con}. Unfortunately,
all these measures are difficult to implement experimentally and
they require substantial efforts to estimate.

An easier way to detect entanglement is using entanglement witnesses
\cite{witness}. Recently it was shown that with non-linear
expressions, that often can be implemented experimentally without
extra effort, any witness can be improved \cite{guhne}. In fact, in
\cite{breuer}, using the universal inverter \cite{posmaps,con}, a
positive map that leads an optimal entanglement witness, in the
sense that it can recognize more entanglement states with positive
partial transpose than any other, is constructed. However
entanglement witness needs to be tailor made for each quantum state.
Hence {\it a priori} knowledge of the state is needed.

Several years ago Schlienz and Mahler proposed a general description
of entanglement using the density matrix formalism \cite{schlienz}.
For the bipartite case they introduce an entanglement tensor who's
components are the covariances between a pair of generators of the
respective algebra for each particle. They show that this tensor is
the difference between the composite density matrix and the tensor
product of the reduced density matrices for each subsystem. By
taking the square form of this tensor one obtains a distance which
is vanishing for any product state and is positive otherwise. This
distance is maximal for maximally entangled states, and it is
invariant under local unitary transformations. The work in
\cite{schlienz} focus on the entanglement of pure states, but they
suggest that it should be possible to extend this result for mixed
states. However, Schlienz and Mahler were ahead of their time
because concurrence had not yet been proposed in 1995, so how to
distinguish between entangled states and a statistical mixture of
separable states is not discussed in \cite{schlienz}.

The use of uncertainty relations in the study of entanglement is
well know for continuous variable \cite{continuos}. In
\cite{hofmann} Hofmann and Takeuchi proposed a generalization the
uncertainty principle to uncertainty sums of local observables for
finite dimensional systems.  They derived local uncertainty
criterion valid for every bipartite separable state. This criterion
was later extended to multiqubit systems and reformulated in a way
that it can be connected with continuous variables thorough the
covariance matrix \cite{guhne1}. Nevertheless, the local uncertainty
sums depends on the sign of the covariances of the local
observables, causing an unnatural asymmetry, and the range of
nonseparable states that this local uncertainty relation are able to
detect is small \cite{hofmann,KH}. The criterion was improved in
\cite{samuelsson}, where with an slightly modification, the
uncertainty relations can detect a larger class of nonseparable
states with the same measurement data as in \cite{hofmann}.

Recently, it was discovered that Schlienz and Mahler's measure
\cite{schlienz} and local uncertainty relations are really just two
sides of the same coin \cite{kothe}. Schlienz and Mahler's measure
can be stated as a criterion (a limit) to ensure the entanglement.
For pure states the measure can be expressed in terms of the
standard concurrence \cite{con}. For a highly entangled state this
measure can even quantify entanglement to some degree. Recently the
measure was tested experimentally \cite{Wang}. In Sec. \ref{Sec:
Correlation measure} we extend the work of Schlienz and Mahler and
Kothe and Bj\"{o}rk on the separability limit and on the relation
between the measure and entanglement invariants.

When trying to detect or quantify entanglement experimentally one
needs to consider that quantum mechanics is based on probabilities.
Hence, in order to obtain as much information as possible when
measuring a quantum state not only a complete set of linearly
independent measures are needed, but they should also optimize the
process. Wootters and Fields \cite{wootfields} showed that
measurements in mutually unbiased basis (MUB) provide a minimal and
optimal way for a complete determination of a quantum state. The
concept of mutual unbiasedness was introduced by Ivanovi\'c
\cite{ivanovic1} who proved that for prime dimension such basis
exist, by an explicit construction. Some time after this concept was
extended for a power of prime dimensional spaces \cite{powerp}. In
Sec. \ref{Sec: Optimal generators} we combine the ideas introduced
in \cite{schlienz}, and in \cite{kothe} with the idea of optimal
experimental estimation of a state, or, in this case, specifically
estimation of its entanglement.

\section{The correlation measure for arbitrary dimensional bipartite systems}
\label{Sec: Correlation measure}

In this section we extend the work on the correlation measure for
two systems, made in \cite{schlienz,kothe}. Specifically, we take
the entanglement measure proposed in \cite{schlienz}, and use it to
prove a criterion for nonseparability and relate it to two
entanglement invariants. An advantage with the criterion is that it
is experimentally measurable, and it only involves correlations
between local measurements.

Consider two systems $A$ and $B$ of dimensions $N_A$ and $N_B$
respectively, where we, without loss of generality, can assume that
$N_A \leq N_B$. The generalization of the bipartite equation is
straightforward \cite{schlienz},
\begin{equation}
\label{Ggeneral} G=\sum_{k=1}^{N_A^2-1}\sum_{l=1}^{N_B^2-1}\lvert
C({\hat{\lambda}_k^{A}},{\hat{\lambda}_l^{B}})\rvert^2,
\end{equation}
where
\[C(\hat{\lambda}_k^{A},\hat{\lambda}_l^{B})=\langle\hat{\lambda}_k^{A}\otimes\hat{\lambda}_l^{B}\rangle
-\langle\hat{\lambda}_k^{A}\otimes\mathbf{\hat{1}}^{B}\rangle\langle\mathbf{\hat{1}}^{A}\otimes\hat{\lambda}_l^{B}\rangle
\]
is the covariance between $\hat{\lambda}_k^{A}$ and
$\hat{\lambda}_l^{B}$ and where $\hat{\lambda}_{k(l)}^{A(B)}$,
$k,l=1,...,N_{A(B)}^2-1$ are the generators of the {\it
su}$(N_{A(B)})$ algebra. They fulfill the relations \beq
\textrm{Tr}(\hat\lambda_k)=0, \quad
\textrm{Tr}(\hat\lambda_k\hat\lambda_l)=\delta_{kl}. \label{genprop}
\eeq In two and tree dimensions a representation of these operators
are the Pauli and the Gell-Mann matrices, respectively, that are
listed in, e.g., \cite{patera}. For higher dimensions an explicit
construction algorithm can be found in \cite{eberly}. Note, however,
that from an experimental point of view, some representations of
{\it su}$(N)$ groups are preferable over others. We will return to
this point in Sec. \ref{Sec: Optimal generators}.

As was pointed out in \cite{schlienz}, and later in \cite{altafini}
for the qubit case, the measure $G$ is proportional the square of
the Hilbert-Schmidt distance between the composite density matrix
and the tensor product of the reduced density matrices,
\begin{equation}
\label{ghs}G=\textrm{Tr}\left
\{(\hat{\rho}-\hat{\rho}^{A}\otimes\hat{\rho}^{B})^2 \right \},
\end{equation}
where $\hat{\rho}^{A(B)}$ is the reduced density matrix for the
subsystem $A(B)$, and $\hat{\rho}$ is that of the composite system.
The density matrices for each subsystem can be written in terms any
set of {\it su}$(N)$ generators (see for example \cite{eberly}),
that is,
\begin{equation}
\label{rho} \hat{\rho}^{A}=\sum_{j=0}^{N_A^2-1}
a_j\hat{\lambda}_j^{A},\qquad \hat{\rho}^{B}=\sum_{j=0}^{N_B^2-1}
b_j\hat{\lambda}_j^{B},
\end{equation}
where here and below, we have taken $\lambda_0^{A(B)} = \identity$
and therefore $a_0 = N_A^{-1}$ and $b_0 = N_B^{-1}$. Since the
direct product of the basis states of the single particles serves as
a basis in the composite system, the density matrix for the total
system can be written as,
\begin{equation}
\label{rhototal} \hat{\rho}=\sum_{k=0}^{N_A^2-1} \sum_{l=0
}^{N_B^2-1} l_{kl}\hat{\lambda}_k^A\otimes \hat{\lambda}_l^B.
\end{equation}
Note that for $k=l=0$, i.e., the first term, we have $l_{00}=(N_A
N_B)^{-1}$ irrespective of $\hat{\rho}$.

The key to probe (\ref{ghs}) is that tracing over one of the
subsystems simply corresponds to choosing the zero component for the
corresponding index \cite{altafini}, in our notation,
\begin{eqnarray*}
a_k&=&\textrm{Tr}(\hat{\rho}^{A}\hat{\lambda}_k^{A})=N_B l_{k0}\\
b_l&=&\textrm{Tr}(\hat{\rho}^{B}\hat{\lambda}_l^{B})=N_A l_{0l}.
\end{eqnarray*}

The measure given by (\ref{Ggeneral}), or equivalently (\ref{ghs})
has some desirable features. One is that in the Hilbert-Schmidt
distance form (\ref{ghs}) is easy to manipulate theoretically. It is
straightforward to see some important properties such that it is
invariant under local unitary transformations. It is also quite
obviously zero for pure, separable states. For the maximally
entangled states, i.e.,
\begin{equation}
\label{maxent}\vert\psi\rangle=\frac{1}{\sqrt{N_A}}\sum_{j=1}^{N_A}
\vert jj \rangle,
\end{equation}
$G$ obtains its maximum, $(N_A^2-1)/N_A^2$.

In order to analyze the properties of the proposed measure, we will
take it in the form (\ref{ghs}). Consider any pure state in the
Schmidt decomposition:
\[\vert\psi\rangle=\sum_{j=1}^{N_A} e^{i\alpha_j}\sqrt{a_j}\vert \psi_j^A \rangle \otimes \vert \psi_j^B \rangle,
\]
where $a_j$, $j=1,\ldots ,N_A$ are real and nonnegative, $a_1 +
\ldots +a_{N_A} = 1$, and $\bra{\psi_i^A}\psi_j^A \rangle =
\bra{\psi_i^B}\psi_j^B \rangle = \delta_{ij}$. Inserting this state
into (\ref{ghs}) one obtains,
\begin{equation}
\label{puregen}G=\sum_{i=1}^{N_A} a_i^4+2\sum_{\substack {i,j=1 \\
i<j}}^{N_A} a_i^2a_j^2-2\sum_{i=1}^{N_A} a_i^3+\sum_{i=1}^{N_A}
a_i^2+2\sum_{\substack{i,j=1\\i<j}}^{N_A} a_ia_j.
\end{equation}

Now consider the generalization of the concurrence for bipartite
systems in higher dimensions
 \cite{concurrence}, the so called $I$-concurrence $\icon$, given by
 \begin{equation}
 \label{concurrence}\icon^2=1-\textrm{Tr}\left \{(\hat{\rho}^{A})^2 \right \}.
 \end{equation}
I-concurrence is an entanglement monotone, that is, it does not
increase on average under local operations and classical
communication. In the $N_A N_B$ dimensional case, the square of the
$I$-concurrence (\ref{concurrence}) reads \cite{concurrence}, \beq
\icon^2 = 1-\sum_{i=1}^{N_A}
a_i^2=2\sum_{\substack{i,j=1\\i<j}}^{N_A} a_i a_j. \eeq
I-concurrence, being only one number, can not make a distinction
between some different kinds of entangled states \cite{badziag,fan}.
That is, states may have the same I-concurrence although they cannot
be transformed one into the other using local operations and
classical communication (LOCC). Nielsen \cite{nielsen} gives
necessary and sufficient conditions for state transformation
processes and in \cite{fan} a concurrence hierarchy is defined. We
know from that work that one needs $N_A-1$ independent invariants
under local unitary transformations in a $N_A$-level quantum system
for a complete characterize of entanglement. In our case,
complementing the concurrence (\ref{concurrence}), we will consider
the 3-concurrence $\threecon$, another invariant under local unitary
transformations that is related with the entanglement between the
superposition-state triads, and which does not increase under LOCC
\cite{fan}, \beq
\threecon=\sum_{\substack{i,j,k=1\\i<j<k}}^Na_ia_ja_k .
\label{c3}\eeq Using (\ref{concurrence}) and (\ref{c3}), we can,
after some algebra, obtain a relation between $G$, $\icon$ and
$\threecon$ for pure states:
\begin{equation}
G=\icon^4+\icon^2-6\threecon. \label{Eq: G function}
\end{equation}
As we can see, the measure $G$ is a function of two of the
invariants of the $N_A-1$ necessary for a complete characterization
of the entanglement \cite{badziag,nielsen,fan}.

Now, we will propose a separability criterion for $N_A
N_B$-dimensional systems.  The limit to ensure entanglement for two
qubits is $G > 1/4$. Note that in Ref. \cite{kothe} the derived
limit is a factor 4 higher because of a different definition (by a
factor of two) of the group generators (\ref{genprop}). We shall
show that this limit is independent of the bipartite system
dimensionality.

A maximally correlated separable state has the form \beq
\hat{\rho}=\sum_{j=1}^{N_A} p_j
\ket{\psi_j^A}\bra{\psi_j^A}\otimes\ket{\psi_j^B}\bra{\psi_j^B},
\eeq where $\bra{\psi_i^A}\psi_j^A \rangle = \bra{\psi_i^B}\psi_j^B
\rangle = \delta_{ij}$. The reason the maximally correlated state
must have this form is that only this form allows, by a proper local
unitary transform, or equivalently, by a properly chosen measurement
basis, to get distinctly correlated measurement outcomes. If
detector $j_A$ ``clicks'', indicating that state $\ket{\psi_j^A}$
was detected, this form guarantees that detector $j_B$ will also
click. Hence, the local measurement outcomes are completely
correlated. Using the method of Lagrange multipliers, it is not hard
to find that the maximal value of $G$ for such state, with the
restraint that all $p_j$ are real, nonnegative, and
$\sum_{j=1}^{N_A} p_j=1$, is 1/4. The state achieving this maximum
has the form \beq \hat{\rho} = \frac{1}{2} \left ( \ket{00}\bra{00}
+ \ket{11}\bra{11}\right ), \label{Eq: Maximally correlated
mixed}\eeq for any $N_A$ and $N_B$, and it is clear that any local
transformations will keep the state as an equal statistical mixture
of two tensor products of locally orthogonal states. Hence, the
criterion \beq G > 1/4 \label{Eq: criterion} \eeq ensures that the
state is nonseparable.

Corresponding to the bipartite qubit case, it is possible to derive
a lower limit of $G$ as a function of the state's I-concurrence and
3-concurrence, and this limit is given by (\ref{Eq: G function}).
This expression provides the lower limit because a pure state has
all its correlations in the entanglement, whereas mixed states can
also have statistical correlations, as shown by the example with an
unentangled state in Eq. (\ref{Eq: Maximally correlated mixed}).

However, in contrast to the bipartite qubit case it is difficult to
derive an upper limit to $G$ as a function of $\icon$ and
$\threecon$ as the latter is undefined for mixed states. We also
lack a systematic way of parameterizing general bipartite states
with a given I-concurrence, and therefore we cannot derive the
function's maximum for a given $\icon$, except that we know that
$G$'s global maximum is $1-N_A^{-2}$ for the state given in
(\ref{maxent}). What is clear from numerical simulations, and which
was shown to hold for the bipartite qubit case, is that when $G$ is
close to its maximal value, the range over which $\icon$ and
$\threecon$ can vary while preserving the value of $G$ is very
small. Hence, for highly entangled states $G$ will pinpoint both
$\icon$ and $\threecon$ through (\ref{Eq: G function}) relatively
well.

The criterion (\ref{Eq: criterion}) is sufficient but not necessary.
A simple example of the latter is the isotropic, two qutrit state
\begin{equation}
\label{werner}\hat{\rho}=\frac{1-\alpha}{9}\hat{\mathbf{1}}+\frac{\alpha}{3}\sum_{m,n=1}^3\vert
mm\rangle\langle nn\vert,~~~~-\frac{1}{8}\leq \alpha \leq 1.
\end{equation}
For this state we obtain $G=8\alpha^2/9$, which implies that for
$\alpha < 3/4\sqrt{2}\approx0.53$ our measure (\ref{Eq: criterion})
cannot say anything about separability, but it is known that for
$\alpha>1/4$ the state (\ref{werner}) is entangled \cite{wernsep}.

\section{Optimal measurement estimates of entanglement}
\label{Sec: Optimal generators}

The measure (\ref{Eq: G function}) yields the same value
irrespective of the set of {\it su}$(N)$ generators one uses,
provided that they fulfil (\ref{genprop}). However, from an
experimental point of view it is desirable that the generators are
unbiased. That is, the generators should be as ``different'' from
each other as possible. In the two-qubit case it is natural to take
the {\it su}$(2)$ generators to be the Pauli matrices. These
generators are all mutually unbiased in that the absolute value of
the scalar product between any eigenvector of one generator and any
eigenvector of any other generator equals $2^{-1/2}$. This is not
true for the Gell-Mann matrices where the corresponding eigenstate
overlap spans between 0 to 1.

Since, starting from a finite ensemble of identically prepared
states, we are interested to measure local correlations as well as
possible, we want to minimize the estimation error due to the
probabilistic nature of quantum measurements. This can be done if we
can construct a set of {\it su}$(N)$ operators that simultaneously
constitute a mutually unbiased basis set MUB \cite{wootfields}.
Unfortunately the constructions of such sets depend on the
dimensionality of the space. The qubit space has already been
discussed, and for odd prime and integer powers of odd and even
prime dimensions, it is possible to find one more MUB than the space
dimension, which is what we need.

Let us start with the qutrit case first, and generalize this later.
For each qutrit \cite{ivanovic1,mubstrit}, there exist $4$ MUB, with
$3$ projectors each,
 $\vert\phi_{i,k}\rangle\langle\phi_{i,k}\vert$, where the
subindex $k=1,\ldots,4$ denotes the basis and the subindex
$i=1,\ldots,3$ denotes the element of the basis.

A common way to construct the MUB is finding $4$ unitary matrices
(one of them is the identity), and then transforming the standard
basis (projectors) with them in order to obtain the $4$ projectors
of the MUB (see for example \cite{ivanovic1,mubstrit,patera}).

In \cite{patera} one can find the $8$ generators of {\it su}$(3)$,
wich as a functions of the MUB projectors
$\hat{\rho}_{i,k}=\vert\phi_{i,k}\rangle\langle\phi_{i,k}\vert$ are
given by
\begin{eqnarray}
L_k & = &
\sqrt{\frac{1}{6}}(2\rho_{1,k}-\rho_{2,k}-\rho_{3,k})\quad\mbox{and} \label{Lij}\\
\tilde{L}_k & = & \sqrt{\frac{1}{2}}(\rho_{3,k}-\rho_{2,k}).
\label{Lij2}
\end{eqnarray}
For convenience we label the operators (\ref{Lij}) and (\ref{Lij2})
\begin{equation*}
\label{Ls}L_1=\hat{\lambda}_1,\:
\tilde{L}_1=\hat{\lambda}_2,\ldots,\tilde{L}_4=\hat{\lambda}_8,
\end{equation*} Is easy to check that these operators are generators of the {\it su}$(3)$
algebra, in other words they fulfill (\ref{genprop}).

In the form (\ref{Lij}) and (\ref{Lij2}) we have eight generators in
terms of 12 projectors, so if we insert this formula in
(\ref{Ggeneral}), it seems like that one should need 144
correlations. This is a chimera since for each basis
\begin{equation}
\label{overcomplete}\sum_{i=1}^3\rho_{i,k}=\mathbf{\hat{1}},~~~~\forall~~
k=1,..4,
\end{equation}
and substituting in (\ref{Lij}) and then in (\ref{Ggeneral}), we
obtain our measure in terms of the MUB projectors' covariances:
\begin{eqnarray}
G & =& \sum_{k,l=1}^4\Big(4
\Big[\sum_{i,j=2}^3C(\rho_{ik}^{A},\rho_{j,l}^{B}) \Big]^2
\nonumber \\
&& -\sum_{i=2}^3\sum_{\substack{i',j '=2\\k'\neq l'}}^3C(\rho_{i,k}^{A},\rho_{j,k}^{B})C(\rho_{i,k'}^{A},\rho_{j,l'}^{B})\nonumber\\
&&-3\sum_{i,i'=2}^3C(\rho_{ik}^{A},\rho_{j2}^{B})C(\rho_{ik'}^{A},\rho_{j3}^{B}) \nonumber \\
&&-3\sum_{i,i'=2}^3C(\rho_{i2}^{A},\rho_{jk}^{B})C(\rho_{i3}^{A},\rho_{jk'}^{B})\Big).\label{gmub}
\end{eqnarray}
As before, this leaves us with sixty four correlations to measure
for two qutrits.

Now let us generalize the result above to dimension $N$, where
$N=2n+1$ is an odd prime. Using the notation introduced in
\cite{patera} the $k$-th group of operators is given by
\begin{eqnarray*}
\label{L}L_{l,k}&=&\frac{1}{\sqrt{2(2n+1)}}(O_k^l+O_k^{2n-l+1}),\qquad
l=1,\ldots,n\\
\label{Ltilde}\tilde{L}_{l,k}&=&\frac{i}{\sqrt{2(2n+1)}}(O_k^{2n-l+1}-O_k^l),\qquad
l=1,\ldots,n,
\end{eqnarray*}
for $k=0,\ldots,2n+1$, where $O_k=(AD^k)$ for $k=0,\ldots,2n$, and
$O_{2n+1}=D$. $A$ is the cyclic permutation matrix and $D$ is the
diagonal matrix which elements are the powers of the $N$-th root of
unity, $\omega=e^{2\pi i/N}$, that is
$D=\textrm{diag}\{1,\omega,\omega^2,\ldots,\omega^{2N}\}$ .

We can use the spectral decomposition to obtain these operators in
terms of the MUB projectors,
\begin{eqnarray}
\label{lijgen}L_{l,k}=\sqrt{\frac{2}{2n+1}}\sum_{j=0}^{2n}\cos{(2\pi lj/(2n+1))}\rho_{j,k}\\
\label{lijgen1}\tilde{L}_{l,k}=\sqrt{\frac{2}{2n+1}}\sum_{j=0}^{2n}\sin{(2\pi
lj/(2n+1))}\rho_{j,k},
\end{eqnarray}
where $\rho_{j,k}$ is the $j$-th eigenprojector of the $k$-th MUB,
with eigenvalue $\omega^j$.

Following the procedure made in for two qutrits, the results
(\ref{Ggeneral}), (\ref{lijgen}) and (\ref{lijgen1}), and the fact
that the projector set is overcomplete, one can construct an
entanglement measure similar to (\ref{gmub}).

For the case when, e.g., $N_A=p^k$ is a power of a prime number, one
can construct the generators of the {\it su}$(N_A)$ algebra in a
similar way, with the unitary matrices given, by example in
\cite{wootfields}, \cite{powerp}. On the other hand, when $N_A$ is a
composite number of at least two different prime numbers, the
corresponding set of mutually unbiased bases are unknown. It is even
not known if one can find $N_A + 1$ mutually unbiased bases. The
evidence at hands is negative, so for such systems the estimation
process is likely to be less efficient.

\section{Conclusions}

In this paper we have extend the work made by Schlienz and Mahler
\cite{schlienz} and Kothe and Bj\"ork \cite{kothe}, taking the
entanglement measure proposed in \cite{schlienz}, to bipartite
states of any dimension. For pure states, it can quantify
entanglement in a certain way, and we derived a relation between
this measure, the I-concurrence and the 3-concurrence (two
entanglement invariants). For mixed states, we established a limit
sufficient, but not necessary, to ensure nonseparability (\ref{Eq:
criterion}).

Taking into account that one can determine in an optimal way all
properties of a state measuring  all combinations of local MUB
eigenstate projections and the identity, we have also given the
measure in terms of MUB eigenprojectors.

\section{Acknowledgements}

This work was supported by the Swedish Foundation for International
Cooperation in Research and Higher Education (STINT), the Swedish
Research Council (VR), and the Swedish Foundation for Strategic
Research (SSF).

\end{document}